\documentstyle[aps,prb]{revtex}

\begin{document}

\input epsf.sty
\twocolumn[\hsize\textwidth\columnwidth\hsize\csname %
@twocolumnfalse\endcsname

\draft

\widetext

\title{Observation of New Incommensurate Magnetic Correlations at the Lower
Critical Concentration for Superconductivity ($x=0.05$) in
La$_{2-x}$Sr$_x$CuO$_4$}

\author{S. Wakimoto and G. Shirane}
\address{Department of Physics, Brookhaven National Laboratory, Upton, NY
11973, USA}
\author{Y. Endoh, K. Hirota, S. Ueki and K. Yamada\footnote{
Present address: Institute for Chemical Research, Kyoto University,
Gokasho, Uji 610-0011, Japan}}
\address{Department of Physics, Tohoku University, Sendai 980-8578, Japan}
\author{R. J. Birgeneau, M. A. Kastner and Y. S. Lee}
\address{Department of Physics, Massachusetts Institute of Technology,
Cambridge, MA 02139, USA}
\author{P. M. Gehring and S. H. Lee\footnote{Also at University of Maryland, 
College Park, MD 20742, USA}}
\address{National Institute of Standards and Technology, NIST Center for
Neutron Research, Gaithersburg MD 20899, USA}

\date{\today}
\maketitle

\vspace{-0.1in}

\begin{abstract}

Neutron-scattering experiments have been performed on lightly-doped
La$_{2-x}$Sr$_{x}$CuO$_{4}$ single crystals in both the insulating
($x=0.03,0.04,0.05$) and superconducting ($x=0.06$) regions.  Elastic
magnetic peaks are observed at low temperatures in all samples with
the maximum peak linewidth occuring at the critical concentration
$x_c=0.05$.  New incommensurate peaks are observed only at $x=0.05$,
the positions of which are rotated by 45$^{\circ}$ in reciprocal space
about $(\pi,\pi)$ from those observed for $x \geq 0.06$ in the
superconducting phase.

\end{abstract}


\phantom{.}
]
\narrowtext

%

\begin{figure}
\centerline{\epsfxsize=2.4in\epsfbox{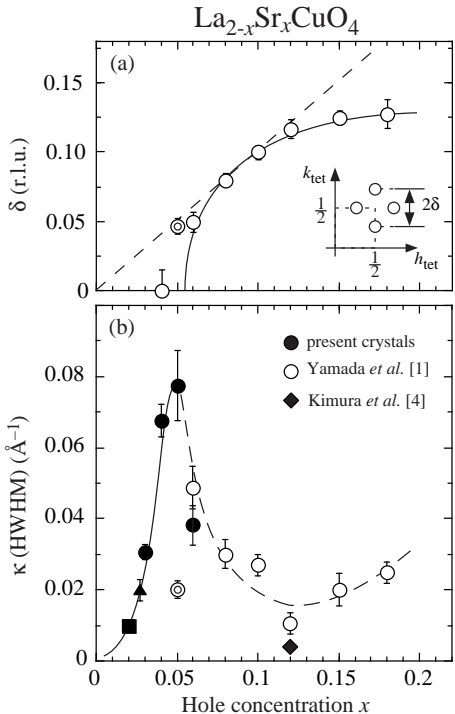}}
\caption{Hole concentration dependence of (a) the incommensurability
$\delta$ of the spin fluctuations, and (b) the neutron-scattering
linewidth of the magnetic IC or C peaks around $(\frac{1}{2},\
\frac{1}{2}, \ -0.3)$. In (a), the dashed line corresponds to
$\delta=x$, whereas the solid line is a guide to the eye.  The inset
shows the IC peak positions in the superconducting phase. 
For $x=0.05$, the IC peak positions are rotated, as described 
in the text, however $\delta$ still represents the magnitude of 
the IC wave vector from $(\frac{1}{2}, \frac{1}{2})$.
In (b), the
solid and dashed lines are guides to the eye.  In both figures, the
closed symbols represent the elastic
components.\protect\cite{M.A.Kastner_98,B.Keimer_92} Open circles
correspond to the widths of 2 to 3.5 meV excitations at
$T=T_{c}$.\protect\cite{K.Yamada_98}.  The double circles at $x=0.05$
denote the new IC elastic peaks.}
\label{Fig:Kappa-Delta}
\end{figure}

The interplay between magnetism and superconductivity has been a
central issue in research on high-$T_{c}$ superconductivity for over a
decade.  Recently Yamada {\it et al.} \cite{K.Yamada_98} carried out a
systematic series of neutron-scattering experiments on
La$_{2-x}$Sr$_{x}$CuO$_{4}$ (LSCO) over a wide range of Sr
compositions to study the evolution of the dynamical spin fluctuations
in the presence of a varying hole concentration.  In these experiments
the fluctuations appear as four incommensurate (IC) inelastic magnetic
peaks centered on the reciprocal lattice position $(\pi,\pi)$ (square
lattice notation, unit lattice constant).  Indexed on a tetragonal
unit cell, the positions of the four IC peaks are
$(\frac{1}{2}\pm\delta,\frac{1}{2})$, and
$(\frac{1}{2},\frac{1}{2}\pm\delta)$ \cite{S.W.Cheong_91}.  (See
Fig.~\ref{Fig:Kappa-Delta}(a)).  The results of Yamada {\it et al.}
\cite{K.Yamada_98} have clarified how the incommensurability $\delta$ changes
with hole concentration $x$ after the onset of superconductivity above
the critical concentration $x_{c}\sim 0.05$ shown in
Fig.~\ref{Fig:Kappa-Delta}(a).  Motivated by the pioneering work of
Tranquada {\it et al.} \cite{J.M.Tranquada_96} in which elastic
magnetic peaks were observed in Nd-doped LSCO, Suzuki {\it et al.} and
Kimura {\it et al.}\cite{T.Suzuki_98} performed neutron-scattering
measurements on La$_{1.88}$Sr$_{0.12}$CuO$_{4}$ ($T_{c} (onset)
=31$~K) and also found sharp elastic magnetic peaks at these same IC
positions with the magnetic transition temperature T$_{m}$ equal to
T$_{c}$.  Kimura {\it et al.}\cite{T.Suzuki_98} also found that the
sharp elastic IC peaks were very weak in a sample with $x=0.10$ and
not observable at all in an optimally doped sample with $x=0.15$,
implying that the magnetic long range order exists only in a very
narrow concentration range near $x=0.12$.  Analogous behavior has been
seen in the system La$_{2-x}$Sr$_x$NiO$_{4+y}$.  In that case, for
hole concentrations n$_h = x + 2y \geq 0.11$, dynamic incommensurate
peaks are also observed, albeit at the positions $(\frac {1} {2} \pm
\frac {\epsilon} {2}, \frac {1} {2} \pm \frac {\epsilon} {2})$, that
is, rotated by 45$^\circ$ with respect to those in
La$_{2-x}$Sr${_x}$CuO$_{4+y}$.  Static order, either short or long
range, is also observed at these positions at low temperatures in the
nickelates.\cite{tranquada_96} Notably, the Ni system is insulating in
both the commensurate and incommensurate magnetic
phases.\cite{tranquada_96}

The magnetic properties of lightly-doped ($0.01<x<0.07$) LSCO have
been studied by various techniques~\cite{M.A.Kastner_98}
including neutron scattering \cite{B.J.Sternlieb_90,B.Keimer_92},
$\mu$SR \cite{B.J.Sternlieb_90,Ch.Niedermayer_98}, and conventional
magnetic measurements \cite{F.C.Chou_95,S.Wakimoto_98}. Keimer {\it et
al.} \cite{B.Keimer_92} reported that at temperatures T
$\stackrel{<}{\sim}$ 20K, La$_{1.96}$Sr$_{0.04}$CuO$_{4}$ exhibits a
broad commensurate (C) elastic peak centered at ($\frac{1} {2},\frac
{1} {2}$), which they ascribed to spins freezing into a spin-glass
(SG) phase.  Chou {\it et al.} \cite{F.C.Chou_95} studied the same
sample using a SQUID magnetometer and confirmed that the sample indeed
exhibited all the features expected for a canonical SG transition at
$T_{g}=7.2$~K.  The same SG behavior is observed in the crystals used
in the present study ($x=0.03, 0.04, 0.05$)
\cite{S.Wakimoto_98}.  Niedermayer {\it et al.}
\cite{Ch.Niedermayer_98} performed a systematic $\mu$SR study over a
wide range of $x$ and found a magnetic transition to a SG-like
state for $0.02\leq x \leq 0.10$, which extends well into the
superconducting regime.

What is the relationship between the broad C elastic peak at $x\sim
0.04$ observed by Keimer {\it et al.}\cite{B.Keimer_92} and the sharp
IC elastic peaks observed by Kimura {\it et al.}\cite{T.Suzuki_98}
that appear to exist only in the vicinity of $x=0.12$?  In this paper,
we try to develop a connection between these two important regions by
performing comprehensive neutron-scattering experiments on single
crystals of LSCO with $x=0.03$, 0.04, 0.05, and 0.06, that bracket the
lower critical concentration for superconductivity, $x_{c}\sim 0.05$.
Our most important and surprising result is that in the sample with
$x=0.05$, which is insulating at low temperatures, we observe sharp
elastic magnetic peaks, albeit at the positions $(\frac{1}{2} \pm
\frac{\epsilon} {2}, \frac{1}{2} \pm \frac {\epsilon}{2})$, that is,
like those in insulating
La$_{2-x}$Sr$_x$NiO$_{4+y}$~\cite{tranquada_96} rather than those in 
superconducting
La$_{2-x}$Sr$_x$CuO$_4$~\cite{M.A.Kastner_98}.  Thus,
the commensurate-incommensurate transition in La$_{2-x}$Sr$_x$CuO$_4$,
which coincides with the onset of superconductivity, proceeds via a
more elaborate route than previously assumed.  This manifestly
represents a major challenge for all theories of high-T$_c$
superconductivity.

Single crystals of La$_{2-x}$Sr$_x$CuO$_4$ with $x=0.03$, 0.04, 0.05,
and 0.06, which were typically 6~mm in diameter and 30~mm in length,
were grown by the travelling-solvent floating-zone method.  
The as-grown crystals were annealed in flowing argon to
reduce the excess oxygen.  We performed x-ray and neutron diffraction
measurements on powdered and single crystal samples, respectively.
The results show the presence of no secondary phases and a good mosaic
spread ($0.3 - 0.5^{\circ}$ full width at half maximum, FWHM),
indicating a high sample quality.  Measurements of the lattice
constants \cite{K.Yamada_98} and the structural phase transition
temperatures $T_s$\cite{M.A.Kastner_98} as well as direct iodometric
titration all confirm that the effective $x$ values are essentially
the same as the nominal ones.

The magnetic susceptibilities of the $x=0.03$, 0.04, and 0.05 samples
were fitted to the canonical model of the SG order parameter.  From
this we determined $T_{g}$ to be 6.3, 5.5, and 5.0~K for $x=0.03$,
0.04, and 0.05, respectively.  Combined with the $T_{g}$ values of the
powdered samples \cite{S.Wakimoto_98}, we have constructed a universal
curve for $T_{g}$ vs.\ $x$.  Using this universal curve, we infer that
the actual hole concentration of the La$_{1.96}$Sr$_{0.04}$CuO$_{4}$
single crystal studied by Keimer {\it et al.} \cite{B.Keimer_92} and
Chou {\it et al.} \cite{F.C.Chou_95} is closer to n$_h=0.027$.  In
contrast to the other crystals, the $x=0.06$ sample shows
superconductivity below $T_{c}=12$~K.

The neutron-scattering experiments were carried out on the SPINS cold
neutron triple-axis spectrometer located at the NIST Center for
Neutron Research, and on the HER cold neutron triple-axis spectrometer
located at the JAERI JRR-3M reactor.  Pyrolytic graphite crystals were
used to monochromate and analyze the neutron energies, and a Be filter
was used to remove contamination from higher order neutron energies.
The energy resolution was about 0.25~meV FWHM for both spectrometers.
The crystals were oriented so as to give access to either the $(h\
h\ l)$ or $(h\ k\ 0)$ zone.  Throughout this paper we will index
reflections using the tetragonal $I4/mmm$ crystallographic structure.

La$_{2-x}$Sr$_{x}$CuO$_{4}$ exhibits superconductivity for $x
\stackrel{>}{\sim} 0.06$.  This concentration range is also where
inelastic magnetic peaks appear at the incommensurate reciprocal
lattice positions shown in Fig.~\ref{Fig:Kappa-Delta}(a)
\cite{K.Yamada_98}.  For $x$ less than the critical concentration
$x_{c} \sim 0.05$ an {\em elastic} peak appears at the {\em
commensurate} position $(\frac{1}{2},\frac{1}{2})$ as shown in Fig.\@
2(a) and (b). We plot in Fig.\@ 1(b) the HWHM of the elastic
and inelastic peaks.  These widths are related to, but technically
distinct from, the instantaneous inverse correlation length $\kappa$
which is measured in an energy integrating neutron experiment
\cite{M.A.Kastner_98,B.Keimer_92}.  As may be seen in Fig.\@ 1(b), the
line widths are sharply peaked at $x_c$.

\begin{figure}
\centerline{\epsfxsize=2.4in\epsfbox{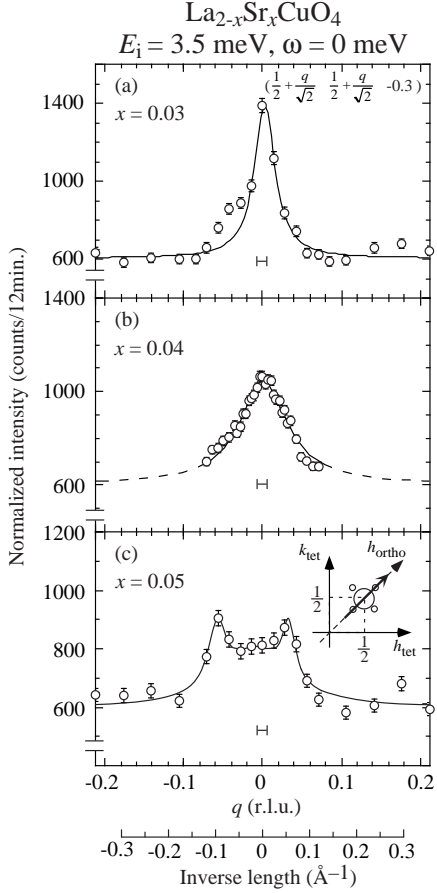}}
\caption{Elastic peak profiles measured around $(\frac{1}{2},\
\frac{1}{2},\ -0.3)$ at $T=2$~K using the SPINS spectrometer.  We use the
$(1/2+q/\protect\sqrt{2},\ 1/2+q/\protect\sqrt{2},\ -0.3)$ notation to
allow a direct comparison of $\delta$ for $x=0.05$ with that for $x
\geq 0.06$ in the tetragonal unit cell.  Solid curves are the results
of least squares fits assuming a single Lorentzian function for
$x=0.03$ and $0.04$, and a triple Lorentzian function (two sharp
incommensurate peaks and one broad commensurate peak) for $x=0.05$.
The small horizontal bars indicate the instrumental resolution
$q$-width.}
\label{Fig:HHL}
\end{figure}

To examine the elastic peaks below $x_{c}$ in more detail, we measured
the peak profile for each crystal in the $(h\ h\ l)$ zone.  As shown
in Fig.~\ref{Fig:HHL}, the C peaks at $T=2$~K for $x=0.03$ and 0.04
are reasonably well fitted with a single Lorentzian convoluted with
the instrumental resolution function.  For the $x=0.03$ sample there
is a small shoulder visible on the low-Q side due to the orthorhombic
splitting.  We also confirmed that the C peak for $x=0.04$ shows only
a weak $l$ dependence, indicating a predominantly two-dimensional (2D)
magnetic character.  To our surprise, however, the peak profile
suddenly changes at $x=0.05$.  Two new IC peaks appear in addition to
the C one, all of which can be fitted to a sum of three Lorentzians.
As may be inferred from the scan trajectory shown in the inset of \linebreak

\noindent 
Fig.~\ref{Fig:HHL}(c), these new IC elastic peaks appear to be rotated
by 45$^{\circ}$ in reciprocal space about $(\frac{1}{2},\frac{1}{2})$
from those in La$_{1.88}$Sr$_{0.12}$CuO$_{4}$.  Moreover,
they resemble the elastic peaks found in
La$_{2-x}$Sr$_{x}$NiO$_{4+y}$.\cite{tranquada_96} None of the IC
peaks, nor the C peak, shows any significant $l$ dependence for
$x=0.05$, that is, the scattering is purely 2D.

\begin{figure}
\centerline{\epsfxsize=2.4in\epsfbox{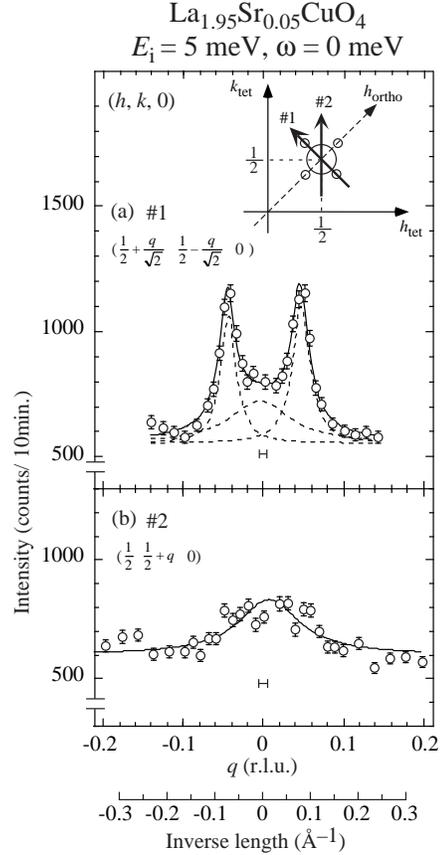}}
\vspace{2mm}
\caption{Elastic peak profiles around $(\frac{1}{2},\frac{1}{2})$ in the
$(h\ k\ 0)$ zone measured in the $x=0.05$ sample at $T=2$~K using the HER
spectrometer.  In (a), we use the $(1/2+q/\protect\sqrt{2},\
1/2-q/\protect\sqrt{2},\ 0)$ notation.  
Solid lines are the results of least squares fits assuming
a triple Lorentzian function for (a), and a
single Lorentzian function for (b).  Dashed lines
in (a) show the individual peak components of the triple Lorentzian
function.  The small bars indicate the instrumental resolution
$q$-width.}
\label{Fig:HK0}
\end{figure}

To verify the reciprocal space positions of these new IC peaks, we
remounted the $x=0.05$ sample in the $(h\ k\ 0)$ zone.  In this case,
the long axis of the resolution ellipse is along the $l$ direction,
thence giving a cleaner signal
from the incommensurate peaks.  As is
clearly shown in Fig.~\ref{Fig:HK0}, the elastic signal consists of
one broad C peak and two sharp IC peaks, the positions of which are
indeed rotated by 45$^{\circ}$ about $(\frac{1}{2},\frac{1}{2})$.  The
peaks are well fitted using a superposition of three Lorentzians for
scan \#1, and a single Lorentzian for scan \#2.  Here we define  \linebreak

\noindent 
the incommensurability $\delta$ for $x=0.05$ as the distance from the IC
peak position to $(\frac{1}{2},\frac{1}{2})$ in reciprocal lattice
units (r.l.u) using the tetragonal notation to maintain consistency
with the prior definition of $\delta$ for $x \stackrel{>}{\sim} 0.06$
in La$_{2-x}$Sr$_x$CuO$_{4+y}$. The results so-obtained are plotted in
Fig.~\ref{Fig:Kappa-Delta}.  The presence of a commensurate peak in
the $x=0.05$ sample may be due to a slight inhomogeneity in the Sr
concentration.  We also carried out some brief elastic measurements on
the $x=0.06$ crystal.  We find weak and somewhat broad elastic
scattering at low temperatures at the q-positions which are identical
to those shown for the inelastic scattering in Figure 1(a).  Thus, the
new IC peaks observed in the $x=0.05$ sample are confined to a very
narrow range in Sr concentrations, just at the insulator to
superconductor transition.

\begin{figure}
\centerline{\epsfxsize=2.4in\epsfbox{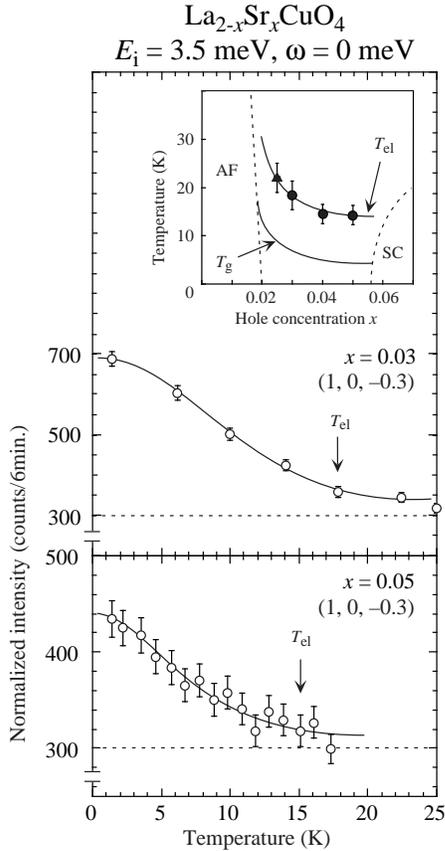}}
\vspace{0mm}
\caption{Temperature dependence of the elastic peak intensity measured
at $(\frac{1}{2},\ \frac{1}{2},\ -0.3)$.  Dashed lines indicate
background levels.  Solid lines are guides to the eye.  Estimated
onset temperatures of the elastic peak are indicated by arrows.
Universal curves for $T_{g}$ and $T_{el}$ are shown in the inset.  The
triangle in the inset represents the datum of Keimer {\it et
al}\protect\cite{B.Keimer_92}.}
\label{Fig:T-dep}
\end{figure}

The temperature dependences of the peak intensities for $x=0.03$, and
0.05 at $(\frac{1}{2},\frac{1}{2}, -0.3)$ are shown in
Fig.~\ref{Fig:T-dep}. Estimates for the onset temperatures $T_{el}$,
where the elastic components first appear, are indicated by arrows.
We have drawn a universal curve for the $x$ dependence of $T_{g}$
using all of the data available at present, and this is shown in the
inset of Fig.~\ref{Fig:T-dep}. The ``$x=0.04$'' sample of Keimer {\it
et al.}\cite{B.Keimer_92} actually sits at $x_{eff}=0.027$ on this
universal curve as well as on the associated curves for the \linebreak

\noindent HWHM (See
Fig.~\ref{Fig:Kappa-Delta}(b)) and $T_{el}$ (See the inset of
Fig.~\ref{Fig:T-dep}) as we already mentioned.

Our results show that fundamental changes in the magnetic properties
of LSCO take place at $x_c=0.05$.  This is most clearly demonstrated
by the sudden appearance of new IC satellite peaks that are rotated by
45$^{\circ}$ with respect to those that exist for $x
\stackrel{>}{\sim} 0.06$ about $(\frac{1}{2},\frac{1}{2})$ at this
critical concentration.  There are two possible interpretations of
this new diffraction pattern.  The first is that
La$_{1.95}$Sr$_{0.05}$CuO$_4$ has a diagonal stripe pattern~\cite{machida} identical
to that in La$_{2-x}$Sr$_x$NiO$_{4+y}$.  In the notation where the
peaks are at $(\frac {1} {2} \pm \frac {\epsilon} {2}, \frac {1} {2}
\pm \frac {\epsilon} {2})$, the measured incommensurability
corresponds to $\epsilon \simeq 0.06 \pm 0.005$,~\cite{notation} close to, but
slightly larger than n$_h=0.05$.  We note that $\epsilon \simeq$ n$_h$
is also observed in La$_{2-x}$Sr$_x$NiO$_{4+y}$ in the incommensurate
phase ($x\geq 0.11$).~\cite{tranquada_96} The second is that the
stripes form a square grid; this will cause the magnetic but not the
charge peaks to rotate by 45$^{\circ}$.  Observations of either the
associated charge peaks or a strong imbalance in the magnetic peak
intensities would choose unambiguously between the diagonal stripe and
grid models.

According to the $\mu$SR study by Niedermayer {\it et al.}
\cite{Ch.Niedermayer_98}, the dependence of $T_{g}$ on hole
concentration is almost the same as that of our crystals (shown in the
inset of Fig.~\ref{Fig:T-dep}).  However, they observed no clear
signals at $T_{el}$.  A possible explanation for this fact is that the
observation time constant of $\mu$SR is much longer than that of
neutron scattering.  Thus, the ``static'' AF correlations reported
here might exist instantaneously and be observable only by neutron
scattering.  If this speculation is correct, the elastic signals must
be {\em quasi-elastic} rather than truly elastic.

We thank V.\ J.\ Emery, Y.\ Fujii, H.\ Fukuyama, S.\ Kawarazaki, P.\
A.\ Lee, K.\ Machida, K.\ Nemoto, M.\ Onodera and J.\ M.\ Tranquada.  The present
work was supported by the US-Japan Cooperative Research Program on
Neutron Scattering.  The work at Tohoku has been supported by a
Grant-in-Aid for Scientific Research of Monbusho and the Core Research
for Evolutional Science and Techonology (CREST) Project sponsored by
the Japan Science and Technology Corporation.  The work at MIT was
supported by the NSF under Grant No.\ DMR97-04532 and by the MRSEC
Program of the National Science Foundation under Award No.\
DMR98-08941.  The work at Brookhaven National Laboratory was carried
out under Contract No.\ DE-AC02-98CH10886, Division of Material
Science, U.\ S.\ Department of Energy.


\begin{references}
\bibitem{notation} Note that $\epsilon$ is larger than $\delta$ plotted in
Fig.~\ref{Fig:Kappa-Delta}(a) by a factor of $\sqrt{2}$.
\end{references}
\end{document}